\documentclass[a4paper,fleqn,usenatbib]{mnras}

\usepackage{newtxtext,newtxmath}

\usepackage[T1]{fontenc}
\usepackage{ae,aecompl}

\usepackage{graphicx}	
\usepackage{amsmath}	
\usepackage{amssymb}	

\makeatletter
\newcommand*{\rom}[1]{\expandafter\@slowromancap\romannumeral #1@}
\makeatother

\newcommand{\kms}{\,\text{km}\,\text{s}^{-1}}
\newcommand{\cc}{\,\text{cm}^{-3}}

\newcommand{\Msun}{\,\text{M}_{\sun}}

\title[AGN jet feedback: GPS and CSS sources]{Relativistic jet feedback II: Relationship to gigahertz peak spectrum and compact steep spectrum radio galaxies}

\author[G.V. Bicknell et al.]{
Geoffrey V. Bicknell$^{1}$\thanks{E-mail: geoff.bicknell@anu.edu.au},
Dipanjan Mukherjee$^{1,2}$,
Alexander Y. Wagner$^{3}$,
\newauthor
Ralph S. Sutherland$^{1}$ and Nicole P.H. Nesvadba$^{4}$,
\\
$^{1}$Australian National University, Research School of Astronomy \& Astrophysics, Cotter Rd. Weston, 
ACT 2611, Australia\\
$^{2}$Dipartimento di Fisica Generale, Universita degli Studi di Torino , Via Pietro Giuria 1, 10125 Torino, Italy\\
$^{3}$Center for Computational Sciences, University of Tsukuba, 1-1-1 Tennodai, Tsukuba, Ibaraki, 305-8577, Japan\\
$^{4}$Institut d'Astrophysique Spatiale, CNRS, Centre Universitaire d'Orsay, Bat. 120$-$121, 91405 Orsay, France
}

\date{Accepted 2018 January 2. Received 2017 December 16; in original form 2017 November 9}

\pubyear{2018}

\begin{document}
\label{firstpage}
\pagerange{\pageref{firstpage}--\pageref{lastpage}}
\maketitle

\begin{abstract}
We propose that Gigahertz Peak Spectrum (GPS) and Compact Steep Spectrum (CSS) radio sources are the signposts of relativistic jet feedback in evolving galaxies. Our simulations of relativistic jets interacting with a warm, inhomogeneous medium, utilize cloud densities and velocity dispersions in the range derived from optical observations, show that free-free absorption can account for the $\sim \rm GHz$ peak frequencies and low frequency power laws inferred from the radio observations. These new computational models replace the power-law model for the free-free optical depth in the \citep{bicknell97a} model by a more fundamental model involving disrupted log-normal distributions of warm gas. One feature of our new models is that at early stages, the low frequency spectrum is steep but progressively flattens as a result of a broader distribution of optical depths, suggesting that the steep low frequency spectra discovered by \citet{callingham17a} may possibly be attributed to young sources. We also investigate the inverse correlation between peak frequency and size and find that the initial location on this correlation is determined by the average density of the warm ISM. The simulated sources track this correlation initially but eventually fall below it, indicating the need for a more extended ISM than presently modelled. GPS and CSS sources can potentially provide new insights into the phenomenon of AGN feedback since their peak frequencies and spectra are indicative of the density, turbulent structure and distribution of gas in the host galaxy.     
\end{abstract}


\section{Introduction}
\label{s:intro}

For some time, GPS and CSS sources have been studied as an interesting and numerically important fraction of the population of extragalactic radio sources. Moreover, it has also been recognised that these sources interact strongly with their environment \citep{gelderman94a,devries99a}. In another field, it has been appreciated that feedback by active galaxies is important in establishing the shape of the galaxy luminosity function and the relationship between the masses of black holes and the mass of the bulge or the velocity dispersion of the host galaxy \citep[][and references therein]{magorrian98a,tremaine02a}. GPS and CSS sources potentially represent one type of AGN feedback -- by relativistic jets in existing or forming galaxies. Hence, understanding the physics of this interaction is not only relevant to the radio galaxy phenomenon but also to galaxy formation and evolution in general.

Previously \citep{wagner11a,wagner11b,wagner12a,wagner13a,mukherjee16a}, we have investigated the role of relativistic jets and winds in providing negative feedback to forming or evolving galaxies. The general idea is that the bubbles driven by the jets quench star formation, either by driving out gas from the centres of the galaxies or by creating so much turbulence that local gravitational collapse is inhibited. There is also the alternative possibility of positive feedback wherein star formation may be enhanced by the compression associated with the jet-driven bubbles \citep[e.g.][]{bicknell00a,gaibler12a,silk13a}.  

When the warm dense gas ($T\lesssim 10^4$K), with which the radio jets interact, is confined to a few core radii of the host galaxy, as it is when the distribution is defined by a quasi-hydrostatic equilibrium between the turbulent pressure gradient and the gravitational field, the typical time for the jet to break free of this clumpy environment is of order $10^6 - 10^7 \> \rm yr$ when the sources are a few kpc in size \citep{mukherjee16a}. These time and length scales further suggest comparison with GPS CSS radio sources. These are believed to be either (a) transient or (b) old and frustrated or (c) young and evolving radio galaxies. In a comprehensive review of GPS and CSS sources, \citet{odea98a} concluded that they are young, and evolving into classical double radio galaxies while noting that their evolution may be \emph{temporarily} frustrated by their host environments. This is consistent with the presence of broad, spatially extended emission lines \citep{gelderman94a,devries99a} resulting from interaction between the radio plasma and the ISM. The models that we present here are based on the notion that GPS and CSS sources are young and temporarily frustrated radio galaxies.

It is apparent, therefore, that a study of GPS and CSS sources potentially provides further insight into the physics of feedback by relativistic jets. In this paper, we concentrate on one possible explanation for the turnover in the spectra of these sources. As we show, this relates strongly to the environment of the host galaxy and the interaction of the jets with it. That is, GPS and CSS sources probe the properties of the warm interstellar medium.

Given that GPS and CSS sources appear to be interacting strongly with the surrounding gas, it is natural to investigate the possibility of free-free absorption (FFA) as the cause of the turnover in the radio spectrum. This was first discussed by \citet{kellermann66a} in the context of the radio source PKS~1934-63. For a uniform, external screen model a free-free absorbed spectrum cuts off abruptly at a critical frequency. However, as \citet{kellermann66a} pointed out, a distribution of optical depths leads to a more gradual turnover. \citet{bicknell97a} further investigated this possibility using an analytic model for the expansion of the radio source and the assumption of a power-law distribution of optical depths. \citet{begelman99a} proposed an alternative ``engulfed cloud'' model in which free-free absorption by photoionized clouds swept up into the expanding radio bubble provide the required optical depth. Begelman also argued that the absorption was unlikely to be the result of synchrotron self-absorption (SSA) on the grounds that it would only be the brightest regions of high magnetic field and/or electron density that would have the highest optical depth. 

More recently, a number of observational papers have supported a FFA interpretation: \citet{kameno05a} presented a detailed analysis of 18 GPS type-1 and type-2 sources. Assuming free-free absorption they calculated the resulting optical depths and showed that they are more asymmetric in the type-1 sources, concluding that this feature is consistent with unified schemes \citep[e.g.][]{barthel89a, barthel94a}. \citet{marr01a} also concluded that the GPS source 0108+388 is free-free absorbed since the bright regions of the source do not exhibit a turnover at a higher frequency than the rest of the source -- an argument similar to that of \citet{begelman99a}.  In another paper, \citet{marr14a} analyzed multi-frequency data on the Compact Symmetric Objects J1324+4048 and J0029+3457, constructing FFA and SSA models for both. For the former source, they again argued that the lack of correlation of the modelled SSA optical depth with bright features in the source indicates FFA by a smooth foreground screen of absorbing gas. They concluded that FFA in the second source is likely but that the case is not as strong. In another investigation, \citet{tingay15a} modelled a combination of Australia Telescope Compact Array (ATCA), Parkes and Murchison Widefield Array (MWA) data on PKS~1718-649 and concluded that the \citet{bicknell97a} FFA model best fitted the spectral data, with the caveat that neither an SSA nor an FFA model explained the variability. \citet{jeyakumar16a}, on the other hand, has modelled the relationship between peak frequency and size of GPS and CSS sources adopting SSA as the absorption mechanism. Nevertheless, it is fair to say that the majority of recent papers favour FFA models.

In view of the observational papers noted above, there is strong motivation for further development of models of GPS and CSS sources, which take advantage of the detailed descriptions of jet-ISM interactions afforded by our three-dimensional simulations. There are several reasons for assessing FFA models in this way: (1) The initial log-normal density distribution is a physically realistic representation of warm gas, which is distributed in the galaxy potential consistently with its velocity dispersion. (2) The simulations capture the details of the jet-ISM interaction such as the ``flood and channel flow'' identified by \citet{sutherland07a} and the simultaneous engulfing of clouds and the shocking of clouds impacted by the expanding bow-shock. This physics defines the density distribution of absorbing clouds. (3) There is no need to involve  additional assumptions and free parameters, such as is incorporated in the power-law distribution of optical depths in the \citet{bicknell97a} model. (4) The simulations and related spectral modelling provides a temporal sequence of the evolution of a GPS/CSS source. (5) Using the distributions of density and temperature provided by our simulations we can investigate the FFA scenario more thoroughly and not only gain insight into the details of the absorption process but also provide additional diagnostics for investigating jet-driven AGN feedback.

Hence, in this paper we present the results of 11 simulations of relativistic jets interacting with inhomogeneous interstellar media. In doing so we concentrate on two major properties of these sources: (1) The turnover of the radio spectrum at frequencies $\sim 100 \> \rm MHz -- 10 \> GHz$ and (2) The inverse correlation of turnover frequency with source size \citep{fanti90a,odea97a}. We begin, in the next section 
(\S\ref{s:simulns}) with a description of the simulations followed by the presentation of the resulting radio spectra (\S\ref{s:spectra}) and the relationship between peak frequency and size (\S\ref{s:nu-size}). We conclude with a discussion in \S\ref{s:discussion}.

\section{Simulations}
\label{s:simulns}

In all simulations we are considering the interaction of a jet with the interstellar medium (ISM) of a spherical galaxy. The ISM consists of a hot tenuous spherical halo and an inhomogeneous medium, which is on average also spherical. The inhomogeneous medium is distributed according to a log-normal distribution with a Kolmogorov distribution in Fourier space. The mean density of the inhomogeneous ISM is defined as a function of radius by an hydrostatic equation with a turbulent pressure. See \citet{sutherland07a} and \citet{mukherjee16a} for details.

The parameters of the simulations are given in Tables~\ref{t:galaxy} and \ref{t:jet_ism}. The gravitational potential and the parameters of the hot halo are, for the most part, the same in all simulations.  Table~\ref{t:galaxy} summarises the parameters describing the potential; Table~\ref{t:jet_ism} summarises the parameters of the inhomogeneous ISM. The only gravitational parameter that is varied is the baryonic core radius, $r_{\rm B}$, and its value for each simulation is also provided in Table~\ref{t:jet_ism}.

\begin{table}
\centering
\begin{tabular}{| l | l | l |}
\hline
\multicolumn{2}{|c|}{Parameters} 	& \multicolumn{1}{|c|}{Value}   \\
\hline
Baryonic (stellar) velocity dispersion  & $\sigma _B$	& 250 km s$^{-1}$	\\
Baryonic (stellar) core radius & $r_{\rm B}$ & 0.4, 1.0 kpc \\
Ratio of dark matter to baryonic core radii & $r_{\rm D}/r_{\rm B}$		& 5	\\
Dark matter velocity dispersion & $\sigma_{\rm D}$ & $500 \> \rm km \> s^{-1}$			\\
Halo Temperature  & $T_{\rm h}$		& $10^7 \> \rm K$  \\
Central hot halo density  & $n_{\rm h,0}$		& $0.5 \> \rm cm^{-3}$ 	\\
\hline
\end{tabular}
\caption{Parameters of the gravitational potential and hot halo common to all simulations.}
\label{t:galaxy}
\end{table}

\begin{table}
\centering
\begin{tabular}{| l | c | c | c | c | c |c|}
\hline 
           &    & \multicolumn{4}{|c|}{Warm clouds} &  \\
\cline{3-6} 
Model & $\log P_{\rm jet}$   & $\sigma_{\rm c}$    & $n_0$         &  Mass     & $T_{\rm floor}$ &
 $r_B$\\
      & $\rm ergs \> s^{-1}$ &  $\rm km \> s^{-1}$ & $\rm cm^{-3}$ & $M_\odot$ & $\rm K$ &
      kpc\\
\hline
A  & 44  & 50   & 400  & $6.46 \times 10^9$  & $10^2$ & 1.0 \\
B  & 44  & 100  & 150  & $2.89 \times 10^9$  & $10^4$ & 1.0 \\
C  & 45  & 50   & 400  & $6.46 \times 10^9$  & $10^2$ & 1.0 \\
D  & 45  & 100  & 150  & $2.89 \times 10^9$  & $10^4$ & 1.0 \\
E  & 45  & 100  & 200  & $2.44 \times 10^9$  & $10^2$ & 1.0 \\
F  & 45  & 100  & 300  & $9.24 \times 10^9$  & $10^4$ & 1.0 \\
G  & 45  & 250  & 400  & $6.61 \times 10^9$  & $10^2$ & 1.0 \\
H  & 45  & 250  & 1000 & $3.47 \times 10^9$  & $10^2$ & 0.4 \\
I  & 46  & 250  & 1000 & $3.47 \times 10^9$  & $10^2$ & 0.4 \\
J  & 46  & 250  & 2000 & $4.76 \times 10^{10}$  & $10^2$ & 1.0 \\
K  & 46  & 300  & 1000 & $1.20 \times 10^{10}$  & $10^2$ & 0.4 \\
\hline
\end{tabular} 
\caption{Parameters of jets and inhomogeneous ISM as well as the baryonic core radius. 
$P_{\rm jet}$ is the jet kinetic power; 
$\sigma_{\rm c}$ is the velocity dispersion of warm clouds; $n_0$ is the (ensemble) mean central number density of the warm clouds; Mass refers to the mass of warm gas; $T_{\rm floor}$ is the floor temperature of the simulation; $r_{\rm B}$ is the baryonic core radius.}
\label{t:jet_ism}
\end{table} 

Simulations, B, D and F, were described in detail in \citet{mukherjee16a}. We have performed new simulations, as listed in Table~\ref{t:jet_ism}, to explore the parameter space further.

Our choice of the range of central warm ISM densities ($n_0$) used in these simulations is guided by the common notion that elliptical galaxies form from the merger(s) of disk galaxies. The additional idea, which we are examining, is that the jets produced by the central black hole prevent the further formation of stars either by clearing the gas or making it strongly turbulent. These processes are mainly envisaged as occurring near the epoch of greatest star formation at redshifts $\sim 2$--3 \citep{silk13a} so that we assume typical densities estimated for star-forming galaxies at those redshifts. \citet{shirazi14a} find electron densities $\sim 400 - 2000 \> \rm cm^{-3}$ between redshifts of 2.17 and 3.10; Sanders et al. find a median electron density of $250\cc$ at $z\sim 2.3$ \citep{shirazi14a}. 

Moreover, radio galaxies and quasars at high redshift are typically found to contain of order $10^8\Msun$ of warm ionized gas and, in some cases, up to $10^9\Msun$ \citep{nesvadba11b, cano-diaz12a, carniani15a, kakkad16a}.  The most powerful high-redshift radio galaxies contain up to $10^{10}\Msun$ with typical electron densities of a few $100\cc$\citep{nesvadba17a}. Our initial densities and gas masses are consistent with these observations, albeit in the higher end of the mass range.  

It was shown in \citet{mukherjee16a} that, before the onset of the jet, the warm clouds settle, in the sense that their velocity dispersion decreases. In all cases, clouds were initially assigned the baryonic velocity dispersion of $250 \kms$; in some cases, we settled the clouds to a lower velocity dispersion. The velocity dispersion shown in Table~\ref{t:jet_ism} is the warm gas velocity dispersion assigned in this way. This is consistent with the range of gas velocity dispersions $ \sim 100-300 \kms$ observed in galaxies at $z \sim 2$ \citep{foerster09a}. 

The cooling from collisionally ionised atoms is evaluated using a tabulated cooling function obtained from the MAPPINGS~V code \citep{sutherland17a}, as also done in \citet{mukherjee16a}. Earlier, it was assumed that ionisation from the central AGN may heat the gas to $10^4$ K and hence a cooling floor of $10^4$ K was applied. In the new runs presented here (see Table~\ref{t:jet_ism}, the cooling floor was set to 100 K. This was done for two reasons. Firstly, keeping the entire ISM warm up to $10^4$ implies an external photoionising source with luminosity comparable to the jet power, which we do not include in our simulations. Secondly the region of the ISM to be photoionised can be evaluated by solving the radiative transfer equation \citep[e.g. see][]{blandhawthorn15a}, which we currently lack in our numerical set up. Hence in order to track the energetics of solely the jet-ISM interaction, we allow the gas to cool to lower temperatures.  We do not include molecular cooling which may dominate below $10^4$K. However such effects occur at time scales smaller than the dynamical time scales of the simulation, and strong cooling from dense regions quickly cools the gas to the cooling floor. 

\section{Spectra}
\label{s:spectra}

GPS and CSS sources are distinguished by their convex spectra, with peak frequencies usually between $\sim 100 \> \rm MHz$ and $\sim 10 \> \rm GHz$. In our simulations, the evolving radio spectra are calculated assuming synchrotron emission from a power-law distribution of relativistic electrons (with electron energy spectral index, $a=2.2$) and free-free absorption by the ambient thermal gas. 

The surface brightness, $I_\nu$, is calculated over a plane perpendicular to the line of sight, which is perpendicular to the jet axis. This is appropriate for GPS and CSS radio galaxies; we are not investigating relativistic beaming effects that would be present in GPS and CSS quasars. Let $j_\nu$ be the synchrotron emissivity and $\alpha_\nu$ the free=free absorption and $s$ the path length through the source. The equation for for the surface brightness is:
\begin{equation}
\frac {dI_\nu}{ds} = j_\nu - \alpha_\nu I_\nu
\end{equation}
Our simulations do not include magnetic field and do not evaluate the fraction of internal energy in relativistic electrons. Hence, we approximate the relativistic electron energy density and magnetic energy density as fixed fractions, $f_{\rm e}$ and $f_{\rm B}$ respectively, of the internal energy density. The details of the expressions for the emissivity and absorption coefficient are described in Appendices \ref{s:spectrum}-\ref{s:power}. 

For the calculations presented here we adopt nominal values of 0.1 for each fraction. However, the results are easily scaled to other values. For example, inverse Compton and synchrotron models of the X-ray and radio emission from a sample of FRII radio galaxy lobes indicate that the magnetic flux density peaks at approximately 0.7 of the equipartition value and that the particle content of the lobes is electron dominated \citep{croston05a} implying that $(f_{\rm e},f_{\rm B}) \approx (0.65,0.35)$. For an electron-positron plasma in equipartition with the magnetic field, $(f_{\rm e},f_{\rm B}) \approx (0.5,0.5)$. In these cases the calculated spectral powers are respectively factors of 62 and 91 times higher as indicated by the dependence of the emissivity $j_\nu$ on $f_{\rm e}$ and $f_{\rm B}$ in equation~\ref{e:jnu}. The spectral \emph{shapes} inferred here are independent of the values of $f_{\rm e}$ and $f_{\rm B}$ since they depend solely on the power-law emissivity and the absorption by thermal gas.

\begin{figure}
	\centering
	\includegraphics[width = \columnwidth]{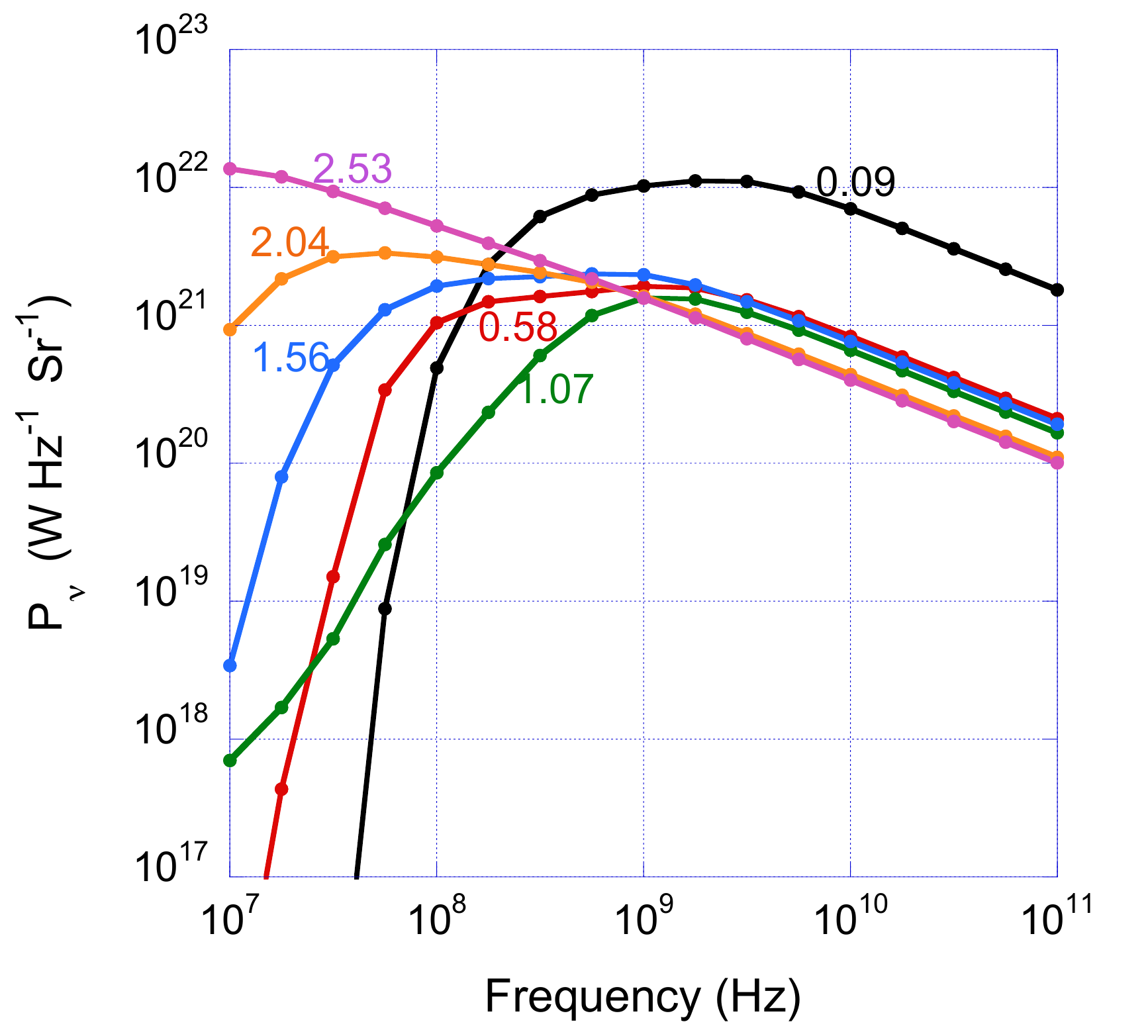}
\caption{Evolution of the radio spectrum for simulation G. Each spectrum is labelled by the time in Myr since the jet was launched.}
\label{f:spectrum}
\end{figure}

As an example, Figure~\ref{f:spectrum} shows the evolution of the radio spectrum from $10^7 - 10^{11} \> \rm Hz$ for simulation G, beginning at an age of 90~kyr. Initially the spectral peak occurs at approximately $2.3 \times 10^9 \> \rm Hz$ and it subsequently moves to lower frequencies as the radio source becomes larger and the average optical depth at a fixed frequency decreases. This evolution is a result of decreasing ambient density and path length through the absorbing gas.  

The distribution of optical depths is determined both by the initial inhomogeneous distribution of dense gas leading to a more variable covering of the source as it evolves in size, and further disruption of this distribution by the emerging jet and lobes. 

\begin{figure}
	\centering
	\includegraphics[width = \columnwidth]{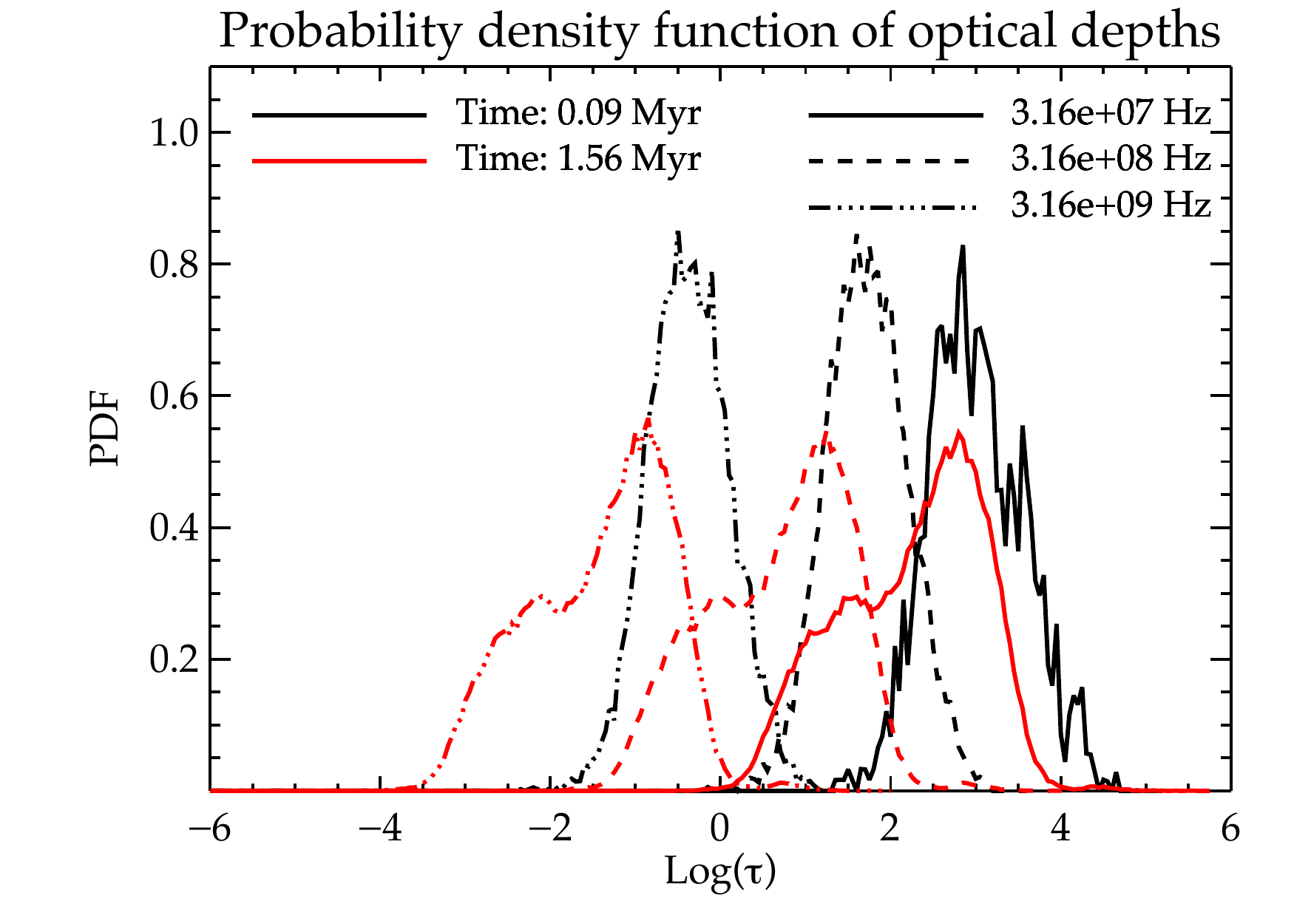}
\caption{Distribution of the free-free optical depth for simulation G at frequencies of $10^{7.5}, 10^{8.5}$ and $10^{9.5} \> \rm Hz$ at times $t=0.09$ and $1.56 \> \rm Myr$ since the jet was launched.}
\label{f:tau}
\end{figure}

The role of the distribution of optical depths is illustrated by Figure~\ref{f:tau} which shows the distribution of optical depths over the radio emitting region at three different frequencies and at times of 0.09 and 1.56 Myr. At 1.56 Myr the average optical depth is lower and the distribution of optical depths is broader than at 0.09~Myr. Note, in particular, the extension of the distribution of the $3.16 \times 10^7 \> \rm Hz$ optical depth back towards $\tau=1$ at $t=1.56 \> \rm Myr$ compared to 
$\tau=100$ at $t=0.09 \> \rm Myr$. This wider distribution of optical depths leads to a power-law low frequency spectrum instead of an abrupt cut-off. In effect, the integrated spectrum consists of the superposition of a number of free-free absorbed spectra, truncated at different frequencies. 
Another feature of the spectral evolution is that the low frequency spectrum is initially steep, reflecting the coverage of the source by an almost uniform optical depth of free-free absorbing gas. At 2.53~Myr, the source is almost a pure power law between $10^7$ and $10^{11} \> \rm Hz$.

\begin{figure*}
\includegraphics[width=\textwidth]{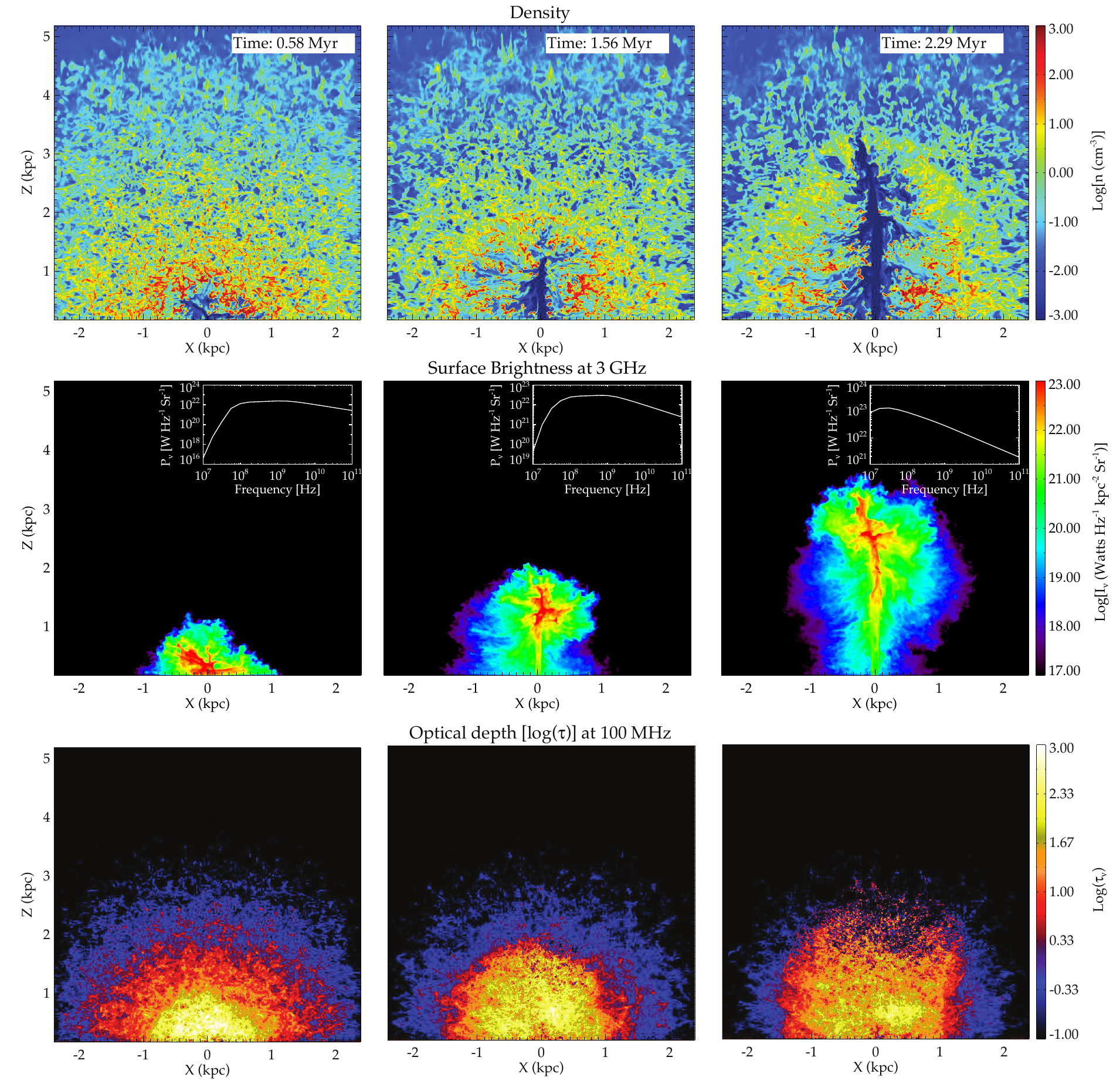}
\caption{Top panels: Mid-plane evolution of density in Simulation G. Second row of panels: Mid-plane evolution of surface brightness at 3~GHz. Third row: Evolution of optical depth through whole volume.}
	\label{f:rhoplots}
\end{figure*}

\begin{figure*}
	\centering
	\includegraphics[width = \textwidth]{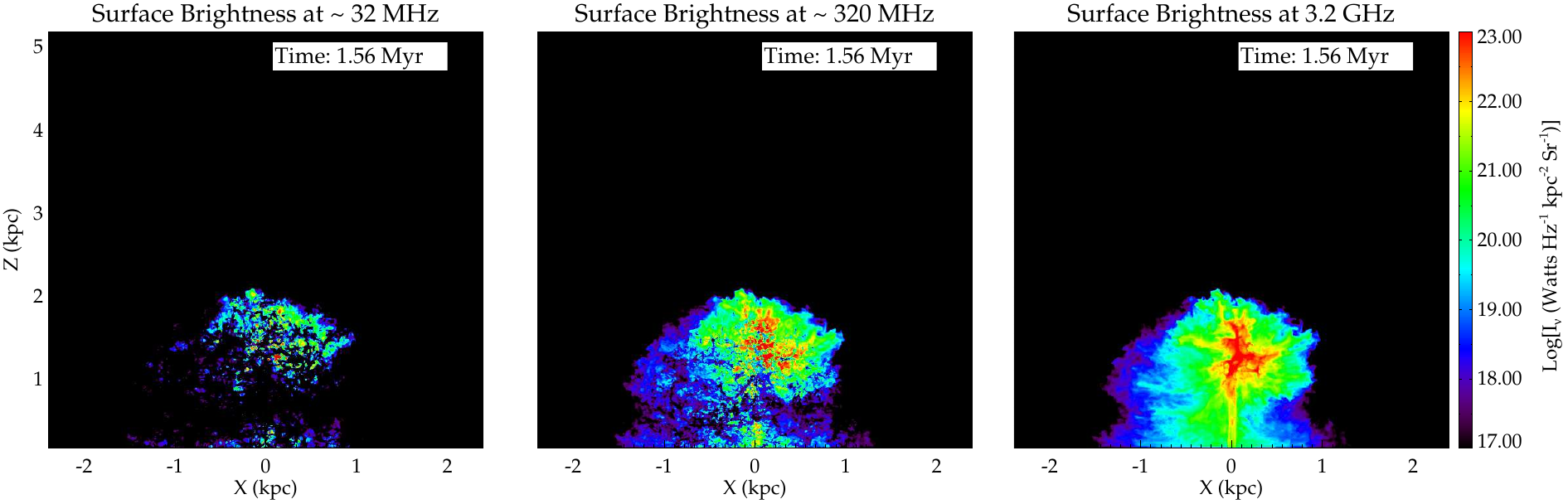} 
	\caption{Surface brightness of Simulation G at a time of 1.56~Myr at 3 frequencies. Left panel: 32~MHz. Middle panel: 320~MHz. Right panel: 3.2~GHz. This figure illustrates the distribution of surface brightness caused by the distribution of optical depths. }
	\label{f:Inuvsfreq}
\end{figure*}

A complementary view of this evolution is shown in Figures~\ref{f:rhoplots} and \ref{f:Inuvsfreq}. The top row of panels of Figure~\ref{f:rhoplots} shows the mid-plane density at 3 separate times of 0.58, 1.56 and 2.29~Myr. The second row of panels shows the corresponding radio surface brightness images at 3~GHz and the related spectra. The third row of panels shows the optical depth throughout the entire volume showing it decreasing, particularly in the central region, as the source evolves. Figure~\ref{f:Inuvsfreq} shows the radio surface brightness at frequencies of 32~MHz, 320~MHz and 3.2 GHz at a time of 1.56~Myr. The effect of increasing absorption with lower frequencies is evident, especially in the central regions of the source.

For those GPS/CSS sources that are observed at an early time in their evolution, the steepness of their spectra predicted by these models serves to distinguish the mechanism for the low frequency turnover from the alternative of synchrotron self absorption. In the latter case, the spectrum cannot be steeper than a slope of 2.5.  Utilising the GLEAM survey \citet{hurley-walker17a}, \citet{callingham17a} have noted fifteen examples of peaked spectra (out of a sample of 1,483) where the low frequency slope is close to or steeper that 2.5, supporting a free-free absorption interpretation for the low frequency turnover in at least those sources.

\begin{figure*}
	\centering
	\includegraphics[width = \textwidth]{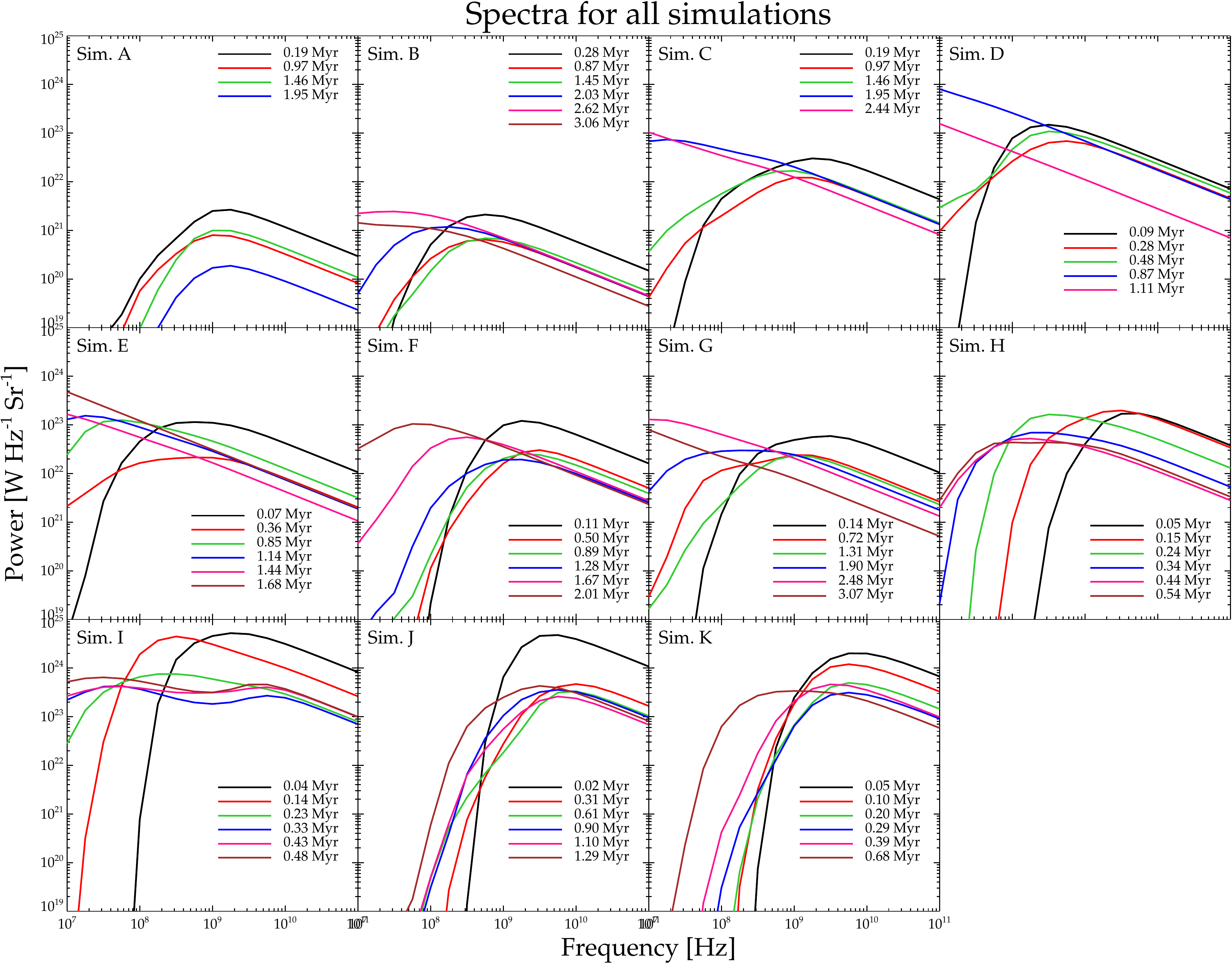}
	\caption{Summary of spectra of all simulations showing the evolution from steep to less steep as each source evolves.}
	\label{fig.spectra_allsim}
\end{figure*}

Figure~\ref{fig.spectra_allsim} displays the evolution of the spectra for all simulations. All spectra show the characteristic of a steep initial spectrum which becomes less steep as the source evolves and in which the peak frequency usually moves to lower frequencies. 

\section{Peak frequency and size}
\label{s:nu-size}

The evolution of the turnover frequency shown in Fig.~\ref{f:spectrum} for simulation~G is typical of most of our simulations. The question then arises as to whether the well-known inverse correlation between turnover frequency and source size \citep{odea97a} is reproduced by these simulations. In particular, following the analytical model of \citet{bicknell97a}, is it feasible that the \citet{odea97a} inverse correlation represents the tracks of evolving GPS/CSS sources? In order to examine this idea, we have determined the turnover frequency and sizes of the simulated radio galaxies as a function of time and over-plotted the evolutionary tracks on the compilation of GPS and CSS peak frequency -- size data in \citet{odea97a}. For the purpose of this comparison, we have selected only radio galaxies and excluded quasars since the spectra of the latter are affected strongly by beaming and their projected sizes are affected by orientation.

\begin{figure*}
	\centering
	\includegraphics[width = \textwidth]{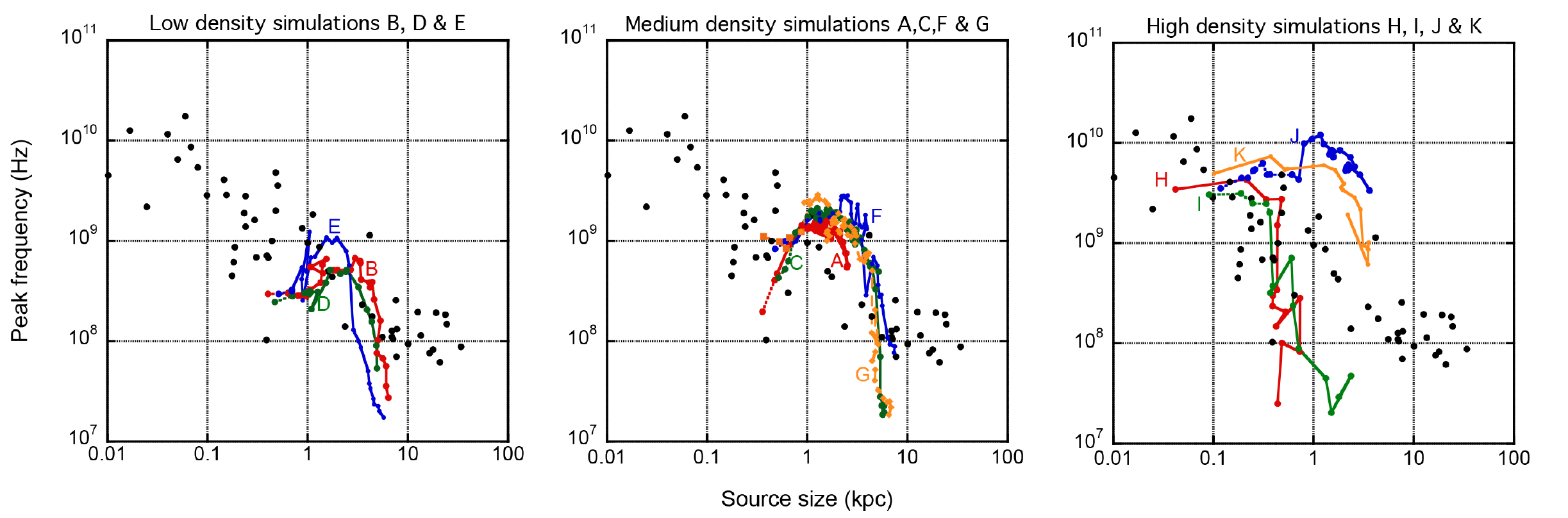}
\caption{Evolutionary tracks of the low density (left panel), medium density (middle panel) and high density (right panel) simulations.}
\label{f:tracks}
\end{figure*}

In order to provide a clear representation of the evolutionary tracks, we have separated the simulations into three classes (low, medium and high density) based on the central density, $n_0$, of the warm ISM. The tracks for the low density simulations B,D, and E ($n_0 = 150-200 \> \rm cm^{-3}$) are shown in the left panel of Figure~\ref{f:tracks}; the medium density simulations A, C, F and G ($n_0 = 300-400 \> \rm cm^{-3}$) are shown in the middle panel; the high density simulations ($n_0 = 1000-2000 \> \rm cm^{-3}$) are shown in the right hand panel.

The low density simulations start within the band of peak frequency -- size points and remain within that band until the source size is about 3-5~kpc. Following that point the tracks drop below the correlation.

Following an initial transient phase the medium density simulations track the upper edge of the data, subsequently track within the data band and then eventually fall below it at a source size of about 5~kpc.

In both of these cases, the decline of the peak frequency below the correlation may be the result of the  density in our models decreasing too rapidly beyond a few core radii. This point is discussed further below.

The high density simulations show a variety of behaviour. First, consider simulation J, which has a central warm density of $2000 \> \rm cm^{-3}$. The evolutionary track of this simulation lies well above the data, reaching a peak frequency above 10~GHz before it starts to track downwards at approximately the same slope as the data. We attribute this to both (a) Too high a central density and (b) Too large a baryonic core radius (1~kpc). Both of these factors lead to a high free-free optical depth which is maintained until the radio source starts to exit the core region. Simulations H and I have the same initial density ($10^3 \> \rm cm^{-3}$) and core radius (0.4~kpc)) but different jet powers ($10^{45}$ and $10^{46} \> \rm ergs \> \rm s^{-1}$, respectively. These track the data reasonably well until about 
0.5~kpc, where, similarly to the low and medium density simulations, they start to fall below the observed correlation. We attribute the same cause to this behaviour, namely too steep a decrease of the density of warm clouds. Simulation K has the same parameters as simulation I except for a larger velocity dispersion of clouds ($300 \> \rm km \> s^{-1}$ compared to $250 \> \rm km \> s^{-1}$ . This maintains a higher cloud density in the core region so that the simulation tends to track above the data.
 
It is apparent from this discussion that the interplay of baryonic core radius, cloud velocity dispersion and cloud density affect the tracks of the simulations. The main parameter which affects the radial slope of the cloud density is the ratio of dark matter velocity dispersion ($\sigma_{\rm D}$)  to cloud velocity dispersion ($\sigma_{\rm c}$).  The asymptotic dependence on radius is 
$n_{\rm c} \propto r^{-2\sigma_{\rm D}^2/\sigma_{\rm c}^2} \approx r^{-8}$ for 
$\sigma_{\rm D} = 500 \> \rm km \> s^{-1}$ and $\sigma_{\rm c} = 250 \> \rm km \> s^{-1}$ The slope is much less ($\loa 2.4$) for $r \loa 5 \> \rm kpc$ and that is the region within which the simulations track the data best. The relationship of galaxy and gas parameters to the evolutionary tracks of the peak frequency will be addressed in detail in future work.

\section{Discussion}
\label{s:discussion}

In this paper we have investigated the potential relationship between feedback by relativistic jets and two key features of GPS and CSS radio galaxies, namely, the turnover in the spectrum at frequencies typically between 100~MHz and 10~GHz and the inverse correlation between the peak radio frequency and the size of the source. We have shown that the broad distribution of optical depths (see Figure~\ref{f:tau}) caused by density fluctuations in the ISM of the host galaxy provides a natural explanation for the slope of the radio spectrum below the peak frequency with the additional prediction that the spectral slope of young sources may be steeper. The observed presence of spectra with a spectral slope steeper than 2.5 \citep{callingham17a} supports our model since slopes this steep slope cannot be produced by synchrotron self absorption. 
 
The sequence of spectra shown in Figures~\ref{f:spectrum} and \ref{fig.spectra_allsim} typically shows the spectral peak evolving towards lower frequency as the source evolves and becomes larger. As a result of the clumpy nature of the ISM the evolution of the peak frequency is not always monotonic. However, as the evolutionary tracks in Figure~\ref{f:tracks} demonstrate, the overall trend is for the peak frequency to decrease as the source gets larger.
   
The inverse correlation between peak frequency and size \citep{odea97a} suggests the possibility of an evolutionary sequence in which radio galaxies evolve from a high density to low density environment, moving the peak frequency to low frequency and increasing in size as they do so. Our simulations suggest that galaxies turn onto this sequence at a point which depends on the warm ISM density. The low and medium density simulations and two of the high density ones track the data for a factor of 5 in size until the peak frequency falls below the observed correlation. As we noted in \S~\ref{s:nu-size} this may be the result of too steep a slope in the density of warm gas. Two of the high density simulations track well above the data. In one case, this is the result of too high a density; in the other it is the result of the radial slope of the density being too flat because of a relatively large core radius 
$\sim 1 \> \rm kpc$. 

In comparison with the \citet{bicknell97a} analytic model our central number densities at a radius of a kpc are similar to the corresponding \citet{bicknell97a} values $\sim 100 \rm cm^{-3}$. The use of an inhomogeneous medium confirms the \citet{bicknell97a} approach of introducing a distribution of optical depths, although the power-law utilised in that paper differs from the disrupted log-normal distribution produced here.  The details of the distribution of the density of warm gas are being further addressed in work in progress. However, the features of the models, which we have presented so far, specifically the ISM density, its turbulent structure and radial distribution, emphasise that the combination of theoretical models with observational data can shed light on these properties in the host galaxies of GPS and CSS sources.

In our radio galaxy model the density of the ambient gas is clearly important and we have been guided by typical densities in star-forming and radio galaxies at redshifts $\sim 2-3$ (see \S~\ref{s:simulns}). In the \citet{odea97a} dataset on the peak frequency--size relation, which we have chosen for comparison, the mean redshift is 0.97 so that we are making the implicit assumption that the conditions in these sources are similar to those at higher redshift. This assumption can only be addressed by further optical and radio observations of these galaxies. Whatever the outcome of such a comparison, it is clear that the radio spectra in GPS and CSS galaxies probe the structure of the warm ISM and this may be relevant to evolving galaxies at high redshift.

Two issues to resolve are what effect the neglect of the magnetic field may have on these results and what would be the effect of the details of the evolution of the relativistic electron distribution. \citet{asahina17a} have made a promising start on the effect of magnetic fields by including an initially straight magnetic field in the ISM, which has a density structure similar to that employed here. Their simulations do no include cooling. \citet{asahina17a} show that the tension force exerted by the magnetic field as it wraps around the clouds, increases the kinetic energy imparted to the clouds by approximately 30\%. Whilst this increase is both interesting and significant, the morphology of the accelerated clouds does not appear to be substantially affected and their simulations do not signal a major departure from the results we have used here. 

We have estimated the relativistic electron density as a fraction ($f_{\rm e}$) the energy density and that is a reasonable approximation when the jet and lobe plasma are not dominated energetically by entrained gas. Nevertheless, the inclusion of magnetic fields in conjunction with evolution of the relativistic electron distribution will enable the rigorous calculation of the Stokes parameters of the radiation field and the distribution of rotation measure. However, the spectra calculated in this paper are informative since the main features on which we have focused, namely the turnover frequency and the low frequency slope, are determined primarily by the cloud structure and not the synchrotron emission, which provides the background emission illuminating the absorbing clouds.

\section*{Acknowledgements}
This research was supported by the Australian Research Council through the Discovery Project, DP140103341 \emph{The Key Role of Black Holes in Galaxy Evolution}. We acknowledge substantial grants of supercomputing time from the National Merit Allocation Scheme and the Australian National University (ANU) Allocation Scheme. We thank the HPC and IT teams at the National Computational Infrastructure (NCI), the ANU, the Pawsey Supercomputing Centre and RSAA for their professional support in carrying out the simulations and the subsequent analysis. We acknowledge a useful referee's report, which assisted us in improving the presentation of this paper.    

\bibliographystyle{mnras}
\bibliography{gvbrefs}

\begin{thebibliography}{}
\makeatletter
\relax
\def\mn@urlcharsother{\let\do\@makeother \do\$\do\&\do\#\do\^\do\_\do\%\do\~}
\def\mn@doi{\begingroup\mn@urlcharsother \@ifnextchar [ {\mn@doi@}
  {\mn@doi@[]}}
\def\mn@doi@[#1]#2{\def\@tempa{#1}\ifx\@tempa\@empty \href
  {http://dx.doi.org/#2} {doi:#2}\else \href {http://dx.doi.org/#2} {#1}\fi
  \endgroup}
\def\mn@eprint#1#2{\mn@eprint@#1:#2::\@nil}
\def\mn@eprint@arXiv#1{\href {http://arxiv.org/abs/#1} {{\tt arXiv:#1}}}
\def\mn@eprint@dblp#1{\href {http://dblp.uni-trier.de/rec/bibtex/#1.xml}
  {dblp:#1}}
\def\mn@eprint@#1:#2:#3:#4\@nil{\def\@tempa {#1}\def\@tempb {#2}\def\@tempc
  {#3}\ifx \@tempc \@empty \let \@tempc \@tempb \let \@tempb \@tempa \fi \ifx
  \@tempb \@empty \def\@tempb {arXiv}\fi \@ifundefined
  {mn@eprint@\@tempb}{\@tempb:\@tempc}{\expandafter \expandafter \csname
  mn@eprint@\@tempb\endcsname \expandafter{\@tempc}}}

\bibitem[\protect\citeauthoryear{{Asahina}, {Nomura}  \& {Ohsuga}}{{Asahina}
  et~al.}{2017}]{asahina17a}
{Asahina} Y.,  {Nomura} M.,   {Ohsuga} K.,  2017, \mn@doi [\apj]
  {10.3847/1538-4357/aa6c5f}, \href
  {http://adsabs.harvard.edu/abs/2017ApJ...840...25A} {840, 25}

\bibitem[\protect\citeauthoryear{Barthel}{Barthel}{1989}]{barthel89a}
Barthel P.~D.,  1989, ApJ, 336, 606

\bibitem[\protect\citeauthoryear{Barthel}{Barthel}{1994}]{barthel94a}
Barthel P.~D.,  1994, in Bicknell G.~V.,  Dopita M.~A.,   Quinn P.~J.,  eds,
  Astronomical Society of the Pacific Conference Series Vol. 54, The First
  Stromlo Symposium: The Physics of Active Galaxies. p.~175

\bibitem[\protect\citeauthoryear{{Begelman}}{{Begelman}}{1999}]{begelman99a}
{Begelman} M.~C.,  1999, in {R{\"o}ttgering} H.~J.~A.,  {Best} P.~N.,
  {Lehnert} M.~D.,  eds, The Most Distant Radio Galaxies. p.~173

\bibitem[\protect\citeauthoryear{Bicknell}{Bicknell}{2013}]{bicknell13a}
Bicknell G.,  2013, Lecture notes on High Energy Astrophysics:
  http://www.mso.anu.edu.au/$\sim$geoff/HEA/

\bibitem[\protect\citeauthoryear{Bicknell, Dopita  \& O'Dea}{Bicknell
  et~al.}{1997}]{bicknell97a}
Bicknell G.~V.,  Dopita M.~A.,   O'Dea C.~P.,  1997, ApJ, 485, 112

\bibitem[\protect\citeauthoryear{{Bicknell}, {Sutherland}, {van Breugel},
  {Dopita}, {Dey}  \& {Miley}}{{Bicknell} et~al.}{2000}]{bicknell00a}
{Bicknell} G.,  {Sutherland} R.,  {van Breugel} W.,  {Dopita} M.,  {Dey} A.,
  {Miley} G.,  2000, ApJL, 540, 678

\bibitem[\protect\citeauthoryear{{Bland-Hawthorn}, {Sutherland}  \&
  {Webster}}{{Bland-Hawthorn} et~al.}{2015}]{blandhawthorn15a}
{Bland-Hawthorn} J.,  {Sutherland} R.,   {Webster} D.,  2015, \mn@doi [\apj]
  {10.1088/0004-637X/807/2/154}, \href
  {http://adsabs.harvard.edu/abs/2015ApJ...807..154B} {807, 154}

\bibitem[\protect\citeauthoryear{{Callingham} et~al.,}{{Callingham}
  et~al.}{2017}]{callingham17a}
{Callingham} J.~R.,  et~al., 2017, \mn@doi [\apj]
  {10.3847/1538-4357/836/2/174}, \href
  {http://adsabs.harvard.edu/abs/2017ApJ...836..174C} {836, 174}

\bibitem[\protect\citeauthoryear{{Cano-D{\'{\i}}az}, {Maiolino}, {Marconi},
  {Netzer}, {Shemmer}  \& {Cresci}}{{Cano-D{\'{\i}}az}
  et~al.}{2012}]{cano-diaz12a}
{Cano-D{\'{\i}}az} M.,  {Maiolino} R.,  {Marconi} A.,  {Netzer} H.,  {Shemmer}
  O.,   {Cresci} G.,  2012, \mn@doi [\aap] {10.1051/0004-6361/201118358}, \href
  {http://adsabs.harvard.edu/abs/2012A%26A...537L...8C} {537, L8}

\bibitem[\protect\citeauthoryear{{Carniani} et~al.,}{{Carniani}
  et~al.}{2015}]{carniani15a}
{Carniani} S.,  et~al., 2015, \mn@doi [\aap] {10.1051/0004-6361/201526557},
  \href {http://esoads.eso.org/abs/2015A%26A...580A.102C} {580, A102}

\bibitem[\protect\citeauthoryear{{Croston}, {Hardcastle}, {Harris}, {Belsole},
  {Birkinshaw}  \& {Worrall}}{{Croston} et~al.}{2005}]{croston05a}
{Croston} J.~H.,  {Hardcastle} M.~J.,  {Harris} D.~E.,  {Belsole} E.,
  {Birkinshaw} M.,   {Worrall} D.~M.,  2005, \mn@doi [ApJ] {10.1086/430170},
  \href {http://adsabs.harvard.edu/abs/2005ApJ...626..733C} {626, 733}

\bibitem[\protect\citeauthoryear{Fanti, Fanti, Schilizzi, Spencer, Rendong,
  Parma, {van~Breugel}  \& Venturi}{Fanti et~al.}{1990}]{fanti90a}
Fanti R.,  Fanti C.,  Schilizzi R.~T.,  Spencer R.~E.,  Rendong N.,  Parma P.,
  {van~Breugel} W. J.~M.,   Venturi T.,  1990, A\&A, 231, 333

\bibitem[\protect\citeauthoryear{{F{\"o}rster Schreiber} et~al.,}{{F{\"o}rster
  Schreiber} et~al.}{2009}]{foerster09a}
{F{\"o}rster Schreiber} N.~M.,  et~al., 2009, \mn@doi [\apj]
  {10.1088/0004-637X/706/2/1364}, \href
  {http://adsabs.harvard.edu/abs/2009ApJ...706.1364F} {706, 1364}

\bibitem[\protect\citeauthoryear{{Gaibler}, {Khochfar}, {Krause}  \&
  {Silk}}{{Gaibler} et~al.}{2012}]{gaibler12a}
{Gaibler} V.,  {Khochfar} S.,  {Krause} M.,   {Silk} J.,  2012, \mn@doi
  [\mnras] {10.1111/j.1365-2966.2012.21479.x}, \href
  {http://adsabs.harvard.edu/abs/2012MNRAS.425..438G} {425, 438}

\bibitem[\protect\citeauthoryear{{Gelderman} \& {Whittle}}{{Gelderman} \&
  {Whittle}}{1994}]{gelderman94a}
{Gelderman} R.,  {Whittle} M.,  1994, \mn@doi [\apjs] {10.1086/191946}, \href
  {http://adsabs.harvard.edu/abs/1994ApJS...91..491G} {91, 491}

\bibitem[\protect\citeauthoryear{{Hurley-Walker} et~al.,}{{Hurley-Walker}
  et~al.}{2017}]{hurley-walker17a}
{Hurley-Walker} N.,  et~al., 2017, \mn@doi [\mnras] {10.1093/mnras/stw2337},
  \href {http://adsabs.harvard.edu/abs/2017MNRAS.464.1146H} {464, 1146}

\bibitem[\protect\citeauthoryear{{Jeyakumar}}{{Jeyakumar}}{2016}]{jeyakumar16a}
{Jeyakumar} S.,  2016, \mn@doi [\mnras] {10.1093/mnras/stw181}, \href
  {http://adsabs.harvard.edu/abs/2016MNRAS.458.3786J} {458, 3786}

\bibitem[\protect\citeauthoryear{{Kakkad} et~al.,}{{Kakkad}
  et~al.}{2016}]{kakkad16a}
{Kakkad} D.,  et~al., 2016, \mn@doi [\aap] {10.1051/0004-6361/201527968}, \href
  {http://esoads.eso.org/abs/2016A%26A...592A.148K} {592, A148}

\bibitem[\protect\citeauthoryear{{Kameno}, {Inoue}, {Wajima}, {Shen}  \&
  {Sawada-Satoh}}{{Kameno} et~al.}{2005}]{kameno05a}
{Kameno} S.,  {Inoue} M.,  {Wajima} K.,  {Shen} Z.-Q.,   {Sawada-Satoh} S.,
  2005, in {Romney} J.,  {Reid} M.,  eds,  Astronomical Society of the Pacific
  Conference Series Vol. 340, Future Directions in High Resolution Astronomy.
  p.~145

\bibitem[\protect\citeauthoryear{{Kellermann}}{{Kellermann}}{1966}]{kellermann66a}
{Kellermann} K.~I.,  1966, \mn@doi [Australian Journal of Physics]
  {10.1071/PH660577}, \href {http://adsabs.harvard.edu/abs/1966AuJPh..19..577K}
  {19, 577}

\bibitem[\protect\citeauthoryear{{Magorrian} et~al.,}{{Magorrian}
  et~al.}{1998}]{magorrian98a}
{Magorrian} J.,  et~al., 1998, AJ, 115, 2285

\bibitem[\protect\citeauthoryear{{Marr}, {Taylor}  \& {Crawford}}{{Marr}
  et~al.}{2001}]{marr01a}
{Marr} J.~M.,  {Taylor} G.~B.,   {Crawford} III F.,  2001, \mn@doi [\apj]
  {10.1086/319729}, \href {http://adsabs.harvard.edu/abs/2001ApJ...550..160M}
  {550, 160}

\bibitem[\protect\citeauthoryear{{Marr}, {Perry}, {Read}, {Taylor}  \&
  {Morris}}{{Marr} et~al.}{2014}]{marr14a}
{Marr} J.~M.,  {Perry} T.~M.,  {Read} J.,  {Taylor} G.~B.,   {Morris} A.~O.,
  2014, \mn@doi [\apj] {10.1088/0004-637X/780/2/178}, \href
  {http://adsabs.harvard.edu/abs/2014ApJ...780..178M} {780, 178}

\bibitem[\protect\citeauthoryear{{Mukherjee}, {Bicknell}, {Sutherland}  \&
  {Wagner}}{{Mukherjee} et~al.}{2016}]{mukherjee16a}
{Mukherjee} D.,  {Bicknell} G.~V.,  {Sutherland} R.,   {Wagner} A.,  2016,
  \mn@doi [\mnras] {10.1093/mnras/stw1368}, \href
  {http://adsabs.harvard.edu/abs/2016MNRAS.461..967M} {461, 967}

\bibitem[\protect\citeauthoryear{{Nesvadba}, {Polletta}, {Lehnert}, {Bergeron},
  {De Breuck}, {Lagache}  \& {Omont}}{{Nesvadba} et~al.}{2011}]{nesvadba11b}
{Nesvadba} N.~P.~H.,  {Polletta} M.,  {Lehnert} M.~D.,  {Bergeron} J.,  {De
  Breuck} C.,  {Lagache} G.,   {Omont} A.,  2011, \mn@doi [\mnras]
  {10.1111/j.1365-2966.2011.18862.x}, \href
  {http://adsabs.harvard.edu/abs/2011MNRAS.415.2359N} {415, 2359}

\bibitem[\protect\citeauthoryear{{Nesvadba}, {De Breuck}, {Lehnert}, {Best}  \&
  {Collet}}{{Nesvadba} et~al.}{2017}]{nesvadba17a}
{Nesvadba} N.~P.~H.,  {De Breuck} C.,  {Lehnert} M.~D.,  {Best} P.~N.,
  {Collet} C.,  2017, \mn@doi [\aap] {10.1051/0004-6361/201528040}, \href
  {http://adsabs.harvard.edu/abs/2017A%26A...599A.123N} {599, A123}

\bibitem[\protect\citeauthoryear{{O'Dea}}{{O'Dea}}{1998}]{odea98a}
{O'Dea} C.~P.,  1998, \mn@doi [\pasp] {10.1086/316162}, \href
  {http://adsabs.harvard.edu/abs/1998PASP..110..493O} {110, 493}

\bibitem[\protect\citeauthoryear{{O'Dea} \& {Baum}}{{O'Dea} \&
  {Baum}}{1997}]{odea97a}
{O'Dea} C.~P.,  {Baum} S.~A.,  1997, AJ, 113, 148

\bibitem[\protect\citeauthoryear{Shirazi, Brinchmann  \& Rahmati}{Shirazi
  et~al.}{2014}]{shirazi14a}
Shirazi M.,  Brinchmann J.,   Rahmati A.,  2014, The Astrophysical Journal,
  787, 120

\bibitem[\protect\citeauthoryear{{Silk}}{{Silk}}{2013}]{silk13a}
{Silk} J.,  2013, \mn@doi [\apj] {10.1088/0004-637X/772/2/112}, \href
  {http://adsabs.harvard.edu/abs/2013ApJ...772..112S} {772, 112}

\bibitem[\protect\citeauthoryear{{Sutherland} \& {Bicknell}}{{Sutherland} \&
  {Bicknell}}{2007}]{sutherland07a}
{Sutherland} R.~S.,  {Bicknell} G.~V.,  2007, \mn@doi [ApJS] {10.1086/520640},
  \href {http://adsabs.harvard.edu/abs/2007ApJS..173...37S} {173, 37}

\bibitem[\protect\citeauthoryear{{Sutherland} \& {Dopita}}{{Sutherland} \&
  {Dopita}}{2017}]{sutherland17a}
{Sutherland} R.~S.,  {Dopita} M.~A.,  2017, \mn@doi [\apjs]
  {10.3847/1538-4365/aa6541}, \href
  {http://adsabs.harvard.edu/abs/2017ApJS..229...34S} {229, 34}

\bibitem[\protect\citeauthoryear{{Tingay} et~al.,}{{Tingay}
  et~al.}{2015}]{tingay15a}
{Tingay} S.~J.,  et~al., 2015, \mn@doi [\aj] {10.1088/0004-6256/149/2/74},
  \href {http://adsabs.harvard.edu/abs/2015AJ....149...74T} {149, 74}

\bibitem[\protect\citeauthoryear{{Tremaine} et~al.,}{{Tremaine}
  et~al.}{2002}]{tremaine02a}
{Tremaine} S.,  et~al., 2002, ApJ, 574, 740

\bibitem[\protect\citeauthoryear{Vries, O'Dea, Baum  \& Barthel}{Vries
  et~al.}{1999}]{devries99a}
Vries W.~D.,  O'Dea C.~P.,  Baum S.~A.,   Barthel P.~D.,  1999, ApJ, 526, 27

\bibitem[\protect\citeauthoryear{{Wagner} \& {Bicknell}}{{Wagner} \&
  {Bicknell}}{2011a}]{wagner11a}
{Wagner} A.~Y.,  {Bicknell} G.~V.,  2011a, \mn@doi [ApJ]
  {10.1088/0004-637X/728/1/29}, \href
  {http://adsabs.harvard.edu/abs/2011ApJ...728...29W} {728, 29}

\bibitem[\protect\citeauthoryear{{Wagner} \& {Bicknell}}{{Wagner} \&
  {Bicknell}}{2011b}]{wagner11b}
{Wagner} A.~Y.,  {Bicknell} G.~V.,  2011b, \mn@doi [ApJ]
  {10.1088/0004-637X/738/1/117}, \href
  {http://adsabs.harvard.edu/abs/2011ApJ...738..117W} {738, 117}

\bibitem[\protect\citeauthoryear{{Wagner}, {Bicknell}  \& {Umemura}}{{Wagner}
  et~al.}{2012}]{wagner12a}
{Wagner} A.~Y.,  {Bicknell} G.~V.,   {Umemura} M.,  2012, \mn@doi [ApJ]
  {10.1088/0004-637X/757/2/136}, \href
  {http://adsabs.harvard.edu/abs/2012ApJ...757..136W} {757, 136}

\bibitem[\protect\citeauthoryear{{Wagner}, {Umemura}  \& {Bicknell}}{{Wagner}
  et~al.}{2013}]{wagner13a}
{Wagner} A.~Y.,  {Umemura} M.,   {Bicknell} G.~V.,  2013, \mn@doi [ApJl]
  {10.1088/2041-8205/763/1/L18}, \href
  {http://adsabs.harvard.edu/abs/2013ApJ...763L..18W} {763, L18}

\bibitem[\protect\citeauthoryear{{Worrall} \& {Birkinshaw}}{{Worrall} \&
  {Birkinshaw}}{2006}]{worrall06a}
{Worrall} D.~M.,  {Birkinshaw} M.,  2006, in {Alloin} D.,  ed.,  Lecture Notes
  in Physics, Berlin Springer Verlag Vol. 693, Physics of Active Galactic
  Nuclei at all Scales. p.~39 (\mn@eprint {} {astro-ph/0410297}),
  \mn@doi{10.1007/3-540-34621-X_2}

\makeatother
\end{thebibliography}

\appendix
\section{Calculation of the radio spectrum.}
\label{s:spectrum}

In \citet{mukherjee16a} we presented log density images of four simulations of jets of powers ranging from $10^{43}$ to $10^{45} \> \rm ergs \> s^{-1}$ interacting with inhomogeneous interstellar environments consisting of $T \sim 10^4 \> \rm K$ gas immersed in hot halos with a central number density of 
$0.5 \> \rm cm^{-3}$, a temperature of $10^7 \> \rm K$ and a gravitational potential described by a baryonic core radius and velocity dispersion of a kpc and $250 \> \rm km \> s^{-1}$ respectively and dark matter core radius and velocity dispersion of 20~kpc and $500 \> \rm km \> s^{-1}$ respectively. (See \citep{sutherland07a} for the description of the double isothermal gravitational potential.) 
We construct models of the radio spectrum as follows. Let $I_\nu$ be the specific intensity of the radio emission, $j_\nu$ the synchrotron emissivity, $\alpha_\nu$ the free-free absorption coefficient and $s$ the path length through the source (starting on the far side at $s=s_1$ and ending at $s=s_2$). The optical depth at a location $s$ along a ray, measured from the far side of the emitting region is
\begin{equation}
\tau_\nu (s) =  \int_{s_1}^s \alpha_\nu(s^\prime) \> ds^\prime 
\end{equation}
Ignoring scattering, the emergent intensity at $s=s_2$ is
\begin{equation}
I_\nu(s_2) = \int_{s_1}^{s_2} j_\nu(s^\prime) \> \exp [-(\tau_\nu(s_2) - \tau_\nu(s^\prime)] \>  ds^\prime
\label{e:Inu}
\end{equation}

\section{Emissivity.}
\label{s:jnu}
We use the following symbols in the ensuing expressions for emission and absorption: $c$ is the speed of light, $k$ is Boltzmann's constant, $e$ the electronic charge, $m_{\rm e}$ the electron mass,
$r_0$ the electron radius, $N(\gamma) =K \gamma^{-a}$ the number density of electrons per unit Lorentz factor, $\gamma$, $B$ the magnetic field, $\Omega_0$ the non-relativistic electron gyrofrequency ($=e B/m_{\rm e}$ in SI units), $\vartheta$ the angle between the magnetic field and the ray to the observer, $\nu^\prime$ the frequency of radio emission in the plasma rest frame, $\nu$ the observed radio frequency, $\theta$ the angle between the direction of an observer located outside the emitting region and the plasma velocity, $\delta$ the Doppler factor of the plasma with respect to , $n_{\rm e}$ the thermal electron density, 
$n_{\rm i}(Z)$ the density of thermal ions of charge $Z e$, and $T$ is the temperature of the thermal gas.

The rest frame emissivity is estimated as follows. The total synchrotron emissivity (see, e.g. \cite{bicknell13a} for the SI expression) can be expressed in the following way, absorbing parameters such as $e^2/\epsilon_0$ into the electron radius. This approach was initiated in \cite{worrall06a}, and provides expressions, which are independent of cgs or SI units. The rest frame  emissivity (primed frame) from a power-law distribution of electrons in a randomly oriented magnetic field is:
\begin{equation}
j^\prime_{\nu^\prime} = C_2(a) \, 
(m_{\rm e} \, c \, r_0) \, K \, \left( \Omega_0 \sin \vartheta \right)^{(a+1)/2} 
\, {\nu^\prime}^{-\alpha}
\end{equation}
where the spectral index $\alpha = (a-1)/2$ and
\begin{equation}C_2(a) = 3^{a/2} 2^{-(a+9)/2} \pi^{-a/2} 
\frac {\Gamma \left( \frac {a+1}{4} \right) \Gamma \left( \frac{a}{4} + \frac{19}{12}\right) 
\Gamma \left(\frac{a}{4} +\frac{1}{12} \right)}
{\Gamma \left( \frac {a+7}{4} \right)} 
\end{equation}

The magnetic field and evolution of the relativistic electron density are not included in our hydrodynamic simulations so that, in order to estimate the synchrotron emissivity, we adopt the ansatz of assuming that the magnetic energy density and relativistic electron energy density are both fixed fractions $f_{\rm B}$ and $f_{\rm e}$ respectively of the total internal  energy density, $\epsilon_{\rm tot}$. Hence, the non-relativistic gyrofrequency $\Omega_0 = (8 \pi r_0/m_{\rm e})^{1/2} \times (f_{\rm B} \epsilon_{\rm tot})^{1/2}$ and the relativistic electron energy density is 
\begin{equation}
\epsilon_{\rm e} = f_{\rm e} \phi_{\rm jet} \epsilon_{\rm tot}
\end{equation}
where $\phi_{\rm jet}$ is the jet tracer variable.
 
The electron density parameter, $K$ is estimated from the relativistic electron energy density, assuming a power-law between minimum and maximum Lorentz factors $\gamma_1$ and 
$\gamma_2$ respectively, so that 
\begin{equation}
K = g(a, \gamma_1, \gamma_2) \, f_{\rm e} \, \frac {\epsilon_{\rm tot}}{m_{\rm e} c^2}
\end{equation}
where
\begin{equation}
g(a, \gamma_1, \gamma_2) = (a-2) \, \gamma_1^{(a-2)} 
\left[ 1 - \left( \frac {\gamma_2}{\gamma_1} \right)^{-(a-2)}\right]^{-1}
\end{equation}
This function appears in minimum energy calculations. For values of $a>2$ and $\gamma_2 \gg \gamma_1$, $g(a, \gamma_1, \gamma_2)$ depends primarily on $\gamma_1$ but also varies slowly with 
$\gamma_1$.

Relativistic beaming effects are incorporated via the Doppler factor 
$\delta = \Gamma^{-1} (1- \beta \cos \theta)^{-1}$ of the plasma relative to an observer located just outside the emitting region and the dependence of the emissivity and observed frequency upon $\delta$ through $\nu = \delta \nu^\prime$ and $j_\nu = \delta^2 j^\prime_{\nu^\prime}$.

With the above assumptions, the random magnetic field emissivity, in the observer's frame, is given by:
\begin{eqnarray}
\langle j_\nu \rangle = & C_2(a) \, \left( \frac{r_0}{c} \right) 
\left( \frac {8 \pi r_0}{m_{\rm e}} \right)^{(a+1)/4} 
g(a,\gamma_1, \gamma_2) \, \delta^{2+\alpha} \phi_{\rm jet} \nonumber\\
& \times  f_{\rm e} \,  f_{\rm B}^{(a+1)/4} \, 
\epsilon_{\rm tot}^{(a+5)/4} 
\nu^{-\alpha} \label{e:jnu}
\end{eqnarray}

\section{Free-free absorption.} 
\label{s:absorption}

The contribution to the absorption coefficient corresponding to ions of charge $Z e$ and for $h\nu\ll kT$ is:
\begin{equation}
\alpha_\nu (Z) \doteq \sqrt { \frac {32 \pi}{27} } c^2 \, r_0^3 \, 
\left( \frac {kT}{m_{\rm e}c^2} \right)^{-3/2} n_{\rm e} \, n_{\rm i}(Z) \, Z^2 \,
g_\nu(T,Z) \> \nu^{-2}
\label{e:alphanu}
\end{equation}
where the Gaunt factor,
\begin{equation}
g_\nu(T,Z) \doteq \frac {\sqrt 3}{2 \pi} 
\left\{ \ln \left[ \frac{8}{\pi^2} \left( \frac{kT}{m_{\rm e}c^2} \right)^3 \frac {c^2}{r_0^2}
\frac {1}{\nu^2 Z^2} \right] -\sqrt {\gamma_E} \right\}
\label{e:gaunt}
\end{equation}
and $\gamma_{\rm E}\approx 0.577$ is Euler's constant.

\section{Emitted power.}
\label{s:power}

We define a plane ${\cal P}$ perpendicular to the observer's direction. Integrating the specific intensity over this plane gives the emitted power in $\rm W \, Hz^{-1} Sr^{-1}$ (using SI units):
\begin{equation}
P_\nu = \int_{\cal P} I_\nu \> dA \quad \rm W \> Hz^{-1} \> Sr^{-1}  
\end{equation}
Computationally, we divide {\cal P} into a large number of cells, with cell size defined by the cell size of the simulation and estimate $P_\nu$ from:
\begin{equation}
P_\nu \approx \sum_i I_{\nu,ij} \, \Delta A_{ij}
\end{equation}
where the indices $ij$ range over all cells in $\cal P$ and $\Delta A_{ij}$ is the area of the $ij$ cell.

\end{document}